\documentclass{article}
\usepackage[T1]{fontenc} 
\usepackage[utf8]{inputenc} 
\usepackage{ismir}
\usepackage{amsmath,cite,url}
\usepackage{graphicx}
\usepackage{color}

\usepackage[super]{nth}
\usepackage[ruled,linesnumbered]{algorithm2e}

\usepackage{booktabs}

\usepackage{colortbl}

\usepackage{subcaption}

\title{Automatic Detection of Cue Points for DJ Mixing}

\threeauthors
{Mickaël Zehren} {Ume\aa{} Universitet \\ {\tt mzehren@cs.umu.se}}
{Marco Alunno} {Universidad EAFIT Medellín \\ {\tt malunno@eafit.edu.co}}
{Paolo Bientinesi} {Ume\aa{} Universitet \\ {\tt pauldj@cs.umu.se}}

\newtheorem{theorem}{Rule}

\begin{document}
\maketitle

\begin{abstract}
  The automatic identification of cue points is a central task in
  applications as diverse as music thumbnailing, mash-ups generation, and DJ mixing.
  Our focus lies in electronic dance music and in specific cue points, the ``switch points'',
  that make it possible to automatically construct transitions among tracks, mimicking what
  professional DJs do.
  We present an approach for the detection of switch
  points that embody a few general rules we established from interviews with professional DJs;
  the implementation of these rules is based on features extraction and novelty analysis.
  The quality of the generated switch points is assessed both by comparing them with a manually annotated
  dataset that we curated, and by evaluating them individually.
  We found that about 96\% of the points generated by our methodology are of good quality for use in a DJ mix.
\end{abstract}

\section{Introduction}
In recent years, there has been a surge in interest in the automatic generation of DJ mixes, that is, uninterrupted
music sequences constructed by partially overlapping music tracks.
In a DJ mix, successive tracks are synchronized
(i.e., tempo-adjusted and beat-matched), possibly overlapped (for a long or short period), and cross-faded.
Intuitively, the ``switch point'' corresponds to the point in
time when the next track in the sequence becomes musically prevalent over the current track (which eventually fades out).
Since this point might affect the listening experience, typically DJs choose it carefully.
In this article, we investigate the possibility---for a specific music genre and mixing style---to identify
switch points automatically.
In particular, we focus on electronic dance music (EDM) and seamless transitions between tracks, by far the most common
type of transitions in sub-genres such as House music and Techno.
This research constitutes an important building block towards the creation of a fully automatic algorithm for DJ mixes.

\begin{figure*}
  \includegraphics[width=\textwidth]{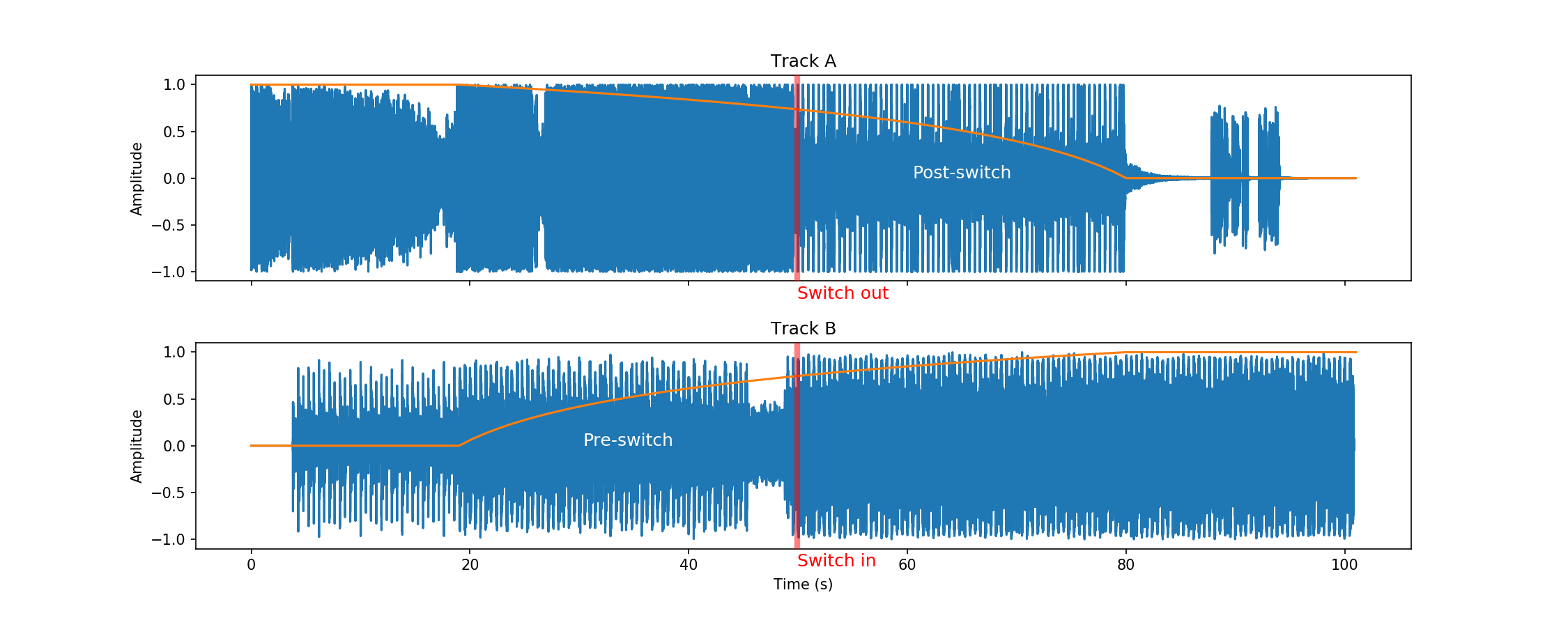}
  \caption{Schematic representation of a simple transition from track A to track B. Orange lines represent volume.} 
  \label{fig:cueDef}
\end{figure*}
Let us consider the scenario of a live performance, where ``track A'' is currently played, and ``track B'' is selected
to be played next. As illustrated in Fig.~\ref{fig:cueDef},
the transition from A to B can be identified by means of three time stamps:
{\sc i}) The point in time when B becomes audible, but A is still dominant---this point marks the beginning of the
``pre-switch'' or ``fade-in'' period;
{\sc ii}) the point in time when B becomes musically prevalent over A---the ``switch'' point (``switch-in'' and ``switch-out'' in relation to track B or A, respectively);
{\sc iii}) the point in time when A becomes inaudible---the end of the ``post-switch'' or ``fade-out'' period.
Depending on the mixing style, the pre- and/or post-switch periods may be instantaneous,
leading to the sudden introduction of track B and/or the sudden removal of track A, respectively.
In the literature, both the start of a fade-in period and the switch points are frequently referred to as ``cue points''.
{\em This study focuses only on switch points, and in particular, on their relative position within track B (``switch-in''), arguably
the most critical position to create a pleasant mix.}
Nonetheless, we believe that our results naturally carry over to fade-in points too.

Our approach for the automatic identification of switch points is inspired by the thought process of professional DJs.
To this end, we conducted semi-structured interviews with DJs specializing in different genres and styles, and collected a set of criteria DJs use to identify viable switch points in a track.
Aiming to follow a knowledge-based approach, we first turned these criteria into high-level rules, and then constructed algorithms to implement them.
We must clarify that these rules are not meant to capture the totality of switch points used by human DJs.
Instead, they provide an algorithm with a sufficient (but not necessary!) set of constrains to automatically identify high-quality switch points.
The effectiveness of our implementation of the rules is evaluated against a dataset that we created---134 tracks manually annotated by experts in music and DJ mixing---and by listening to the generated switch points individually.\footnote{The code is available at https://github.com/MZehren/Automix and the dataset at https://github.com/MZehren/M-DJCUE}
The rest of this article is organized as follows: Sec.~\ref{RelWork} contains a survey of related works.  The rules are presented in Sec.~\ref{Rules}, and the methodology to identify switch points is detailed in Sec.~\ref{System}.  In Sec.~\ref{Evaluation}, results are presented and discussed. Conclusions are drawn in Sec.~\ref{Conclusions}.

\section{Related work} \label{RelWork}
In the last 20 years, several approaches were proposed to automate different stages of a DJ's workflow, and
two criteria emerged to decide where transitions can take place:
“inter-mixability” \cite{gebhardt_psychoacoustic_2016,hirai_musicmixer:_2016,lin_music_2009} and
“intra-mixability” \cite{cliff_hang_2000, bittner_automatic_2017, kim_automatic_2017, vande_veire_raw_2018, schwarz_heuristic_2018, davies_automashupper:_2014, lin_audio_2015}.
The former expresses the compatibility between two tracks (or segments of tracks),
while the latter captures how well-suited a specific position is for a transition,
independently of the next track in the DJ-mix. In our work, we are only concerned with intra-mixability;
inter-mixability will be incorporated in the near future.

Most of the existing techniques to assess intra-mixability rely on structure analysis, a well-defined task in music information retrieval that is typically solved with approaches based on novelty, homogeneity, or repetition~\cite{paulus_audio-based_2010}. In this section, we concentrate on the literature that deals with the identification of switch points for DJ mixes. For this problem, there also exists commercial software such as Mixed In Key\footnote{https://mixedinkey.com} and Pacemaker,\footnote{https://pacemaker.net} one of which we included in our evaluation (Sec.~\ref{Evaluation}).

One of the earliest discussions of switch points identification for the automatic creation of DJ mixes appears in ``Hang the DJ''~\cite{cliff_hang_2000}. There, the idea is to {\sc i}) transition between tracks during segments with no clear pulse (also known as "breakdowns"), and {\sc ii)} identify such segments as the portions of a track in which a beat detection algorithm fails to detect the beat. Unfortunately, this work does not include an evaluation of the results.

In a more recent study~\cite{bittner_automatic_2017}, switch points are collected by combining three different methods: First, crowdsourcing is used to identify drop locations,\footnote{A "drop" is a point of high energy that follows a build up.} then, the structure of the track is determined by a repetition-based algorithm~\cite{mcfee_analyzing_2014}, and finally, downbeat locations are retrieved from Echo Nest,\footnote{http://the.echonest.com} a music intelligence platform~\cite{jehan_creating_2005}. Multiple heuristics are used to prune candidates and select the switch points, and most of the resulting transitions are rated as satisfactory, with only $\approx\!8\%$ of ``bad'' ones.

A recent approach~\cite{kim_automatic_2017} proposes to combine Foote's novelty-based algorithm for structure analysis~\cite{foote_automatic_2000} with deep neural networks; unfortunately, an evaluation of the results is missing.

Besides our current study, both Veire~\cite{vande_veire_raw_2018} and Schwarz et al.~\cite{schwarz_heuristic_2018}
proposed rule-based approaches for the identification of switch points. In his work, Veire aims at the automatic
generation of Drum \& Bass DJ mixes and explains how to improve the quality of the boundaries returned by Foote's
algorithm~\cite{foote_automatic_2000} through rules that encode knowledge of the music genre under consideration. This
method was positively evaluated, but it was not extended to other musical genres.
The goal of Schwarz et al.~is instead ``point-of-sale automatic mixing''; the method improves the structure analysis obtained with the module ``IRCAM SUMMARY''~\cite{kaiser_simple_2013, kaiser_music_2012} by means of heuristics from experts in music branding and the knowledge obtained from a database of 30 tracks manually annotated.

The identification of switch points is a task relevant also for the automatic generation of mash-ups and medleys. For instance, in ``AutoMashUpper''~\cite{davies_automashupper:_2014}, the transitions between tracks are constrained to take place at the boundary between phrases detected with Foote's algorithm.
In medley generation, instead, cuts in vocals can be prevented by extracting phrase structures through voice detection~\cite{lin_audio_2015}.

\section{Rule-based approach} \label{Rules}

By definition, switch points are located at positions in the track where some event of structural importance occurs in the music; therefore, in order to identify them, one has first to detect structural boundaries in the track. This task is simplified, to a certain extent, by the modular nature of the music genre we target, as modularity usually provides structural predictability to music form. Indeed, EDM is highly regular both metrically and formally. Although exceptions exist, these are vastly outweighted by regularity, and should not affect general characterizations:\footnote{Regular features as well as several types of ambiguities of EDM tracks are analyzed in \cite{butler_unlocking_2003}.} EDM is invariably in 4/4, and composed of a periodic repetition of phrases (sometimes called \textit{loops}) as building blocks of its structural elements. It is expected that all structural elements are constituted of two, or a multiple of two, repetitions of a phrase which is itself composed of two, or a multiple of two, bars. In other words, in most EDM tracks every period of four bars anticipates the potential start of a new music section.

Formally, we identify three sections: the intro (the initial section), the core (the central section), and the outro (the last section). Both the intro and the outro may or may not be present. Often, all these sections can be further segmented into portions such as breakdowns (where energy drops), and risers (where energy grows), sometimes followed by a drop (the first beat of a high-energy section) to name a few.
However, not every change in music structure is identified by a DJ as a switch point. Knowledge of the process that leads a DJ to choose where to make a transition would certainly be valuable for an algorithm that is meant to replicate that choice. Therefore, we first encode such knowledge into rules (presented in this section), and then apply them to design an algorithm for the detection of switch points.

The first rule states that structural events are points in time where something new happens on the music surface, and that novelty is interpreted as a boundary between two consecutive music regions. Common parameters in EDM used to convey a sense of change are rhythmic density, loudness, timbre, and harmony.
\begin{theorem}{\bf Novelty.}
  \label{rule:novelty}
  A switch point marks a position of high novelty in rhythmic density, loudness, timbre, and/or harmony.
\end{theorem}

Structural events (hence potential switch points) are not positioned randomly within the rhythmic grid. In EDM, they always coincide with the downbeat of the bar and, due to structural modularity, with the downbeat of the first bar of a period.
\begin{theorem}{\bf Sections detection.}
  \label{rule:switch point position}
  A switch point always occurs on the downbeat at the start of a period.
\end{theorem}

Rule 2 does not lead to the identification of one and only one switch point. For example, it does not solve the ambiguity among multiple viable switch points at the beginning of a track. To make a selection, some of the candidates will be eliminated by the next rule.
\begin{theorem}{\bf Salience detection.}
  \label{rule:switch-point prevalent}
  The section following a switch point has to be able to stand on its own in the mix.
\end{theorem}
We define the ability of ``standing on its own'' as the ``salience'' of a section, that is its property of being constituted of musical elements which are prominent or particularly noticeable. This rule aims at keeping the audience interested. In fact, at the switch, the next track becomes prevalent, and the audience's attention should be caught by any element of novelty and/or interest, such as a new melody, a new bassline or a new drum pattern. All of these events are possible candidates to become switch points.

We repeat that these rules are not intended to model the techniques used by human DJs. For instance, one could argue
that in order to create a sense of anticipation, a DJ wants to switch 4 or 8 bars before the next track exhibits any
element of novelty. We agree. 
However, the role played by what we identify as switch points is uncontested.


\section{Algorithmic approach} \label{System}


Algorithm~\ref{algo:cueDetection} takes a track as input, and returns as output a list of locations that are deemed
viable as switch-in points. A step-by-step description of the procedure follows.

\begin{algorithm}
  \caption{Detection of switch points}
  \label{algo:cueDetection}
  \SetAlgoLined
  \DontPrintSemicolon

  \SetKwFunction{proc}{getSwitchPoints}
  \SetKwProg{myproc}{Procedure}{}{}
  \SetKwComment{Comment}{$\triangleright$\ }{}
  \myproc{\proc{$track$}}{
    $P\leftarrow\{ \}$ \\
    $\{F_{kick}, ..., F_{pcp}\}\leftarrow \text{noveltyFeatures}(track)$ \\
    $B\leftarrow \text{strongBeatDetection}(track)$ \\
    \ForEach{$F \in \{F_{kick}, ..., F_{pcp}\}$}{
      $
        A_{F}\leftarrow
        \begin{cases}
          \text{aggSum}(F, B) & \text{if } F \text{ is sparse} \\
          \text{aggRMS}(F, B) & \text{otherwise}
        \end{cases}
      $ \\
      $N_{F}\leftarrow \text{Ch}() \ast \text{SelfSimilarity}(A_{F})$ \\
      $P_{F}\leftarrow \text{peakPicking}(N_{F})$ \\
      $P=P\cup P_F$
    }
    $P \leftarrow P \cap \text{periodSelect}(\{N_{F_{kick}}, ..., N_{F_{pcp}}\})$ \\
    $P \leftarrow P \cap \text{salienceSelect}(F_{Harmonic})$ \\ 
    \KwRet $P$
  }
\end{algorithm}

\subsection{Feature extraction} \label{Feature}
As Rule \ref{rule:novelty} states, a switch point must be located at a position where at least one of the analyzed music
components (extracted in Line 3 of Alg.~\ref{algo:cueDetection}) shows a high degree of novelty.  To this end, we
empirically found that a combination of seven features (all categorized as “rhythm”, “loudness”, or “spectrum”), yields the best results.

To detect rhythmic changes, we extracted the three main components of the drum set---i.e., kick, snare, and hi-hat---
with Richard Vogl's software~\cite{vogl_drum_2017, vogl_towards_2018}. 
We pay special attention to these components because of their function in stressing segment boundaries in EDM.  For example, the start of the typical 4/4 EDM steady rhythm with the kick drum on each beat, also known as "four to the floor", is a critically important event for the selection of candidates for switch points.

To find loudness variations in a track,
we used Librosa's algorithm {\em harmonic percussive source separation}~\cite{mcfee_librosa:_2015}
to split the signal into its harmonic and percussive components. This proved to be a good approach because it permits to obtain a finer level of detail, if compared to the raw signal, in the energy (and loudness) of both the harmonic and percussive components. This splitting method applies median filtering both across successive frames of the spectrogram to suppress percussive events and across the frequency bins to suppress harmonic components.

We found structural boundaries also by looking at the spectrum of the track. To do so, we used the Constant-Q Transform
(CQT, representing harmony, timbre and loudness) and Pitch Class Profiles (PCP, representing harmony) since they are
known to give good results in music structure analysis~\cite{nieto_systematic_2016}. Both features were extracted with
Librosa by setting the sampling rate at 44.1kHz; the settings for CQT are 84 bins, starting frequency of 32.7Hz, and
hop-length of 512 samples; those for PCP are seven octaves of coverage, and a starting frequency of 27.5Hz.

\begin{figure*}
  a)
  \def\svgwidth{\columnwidth}
  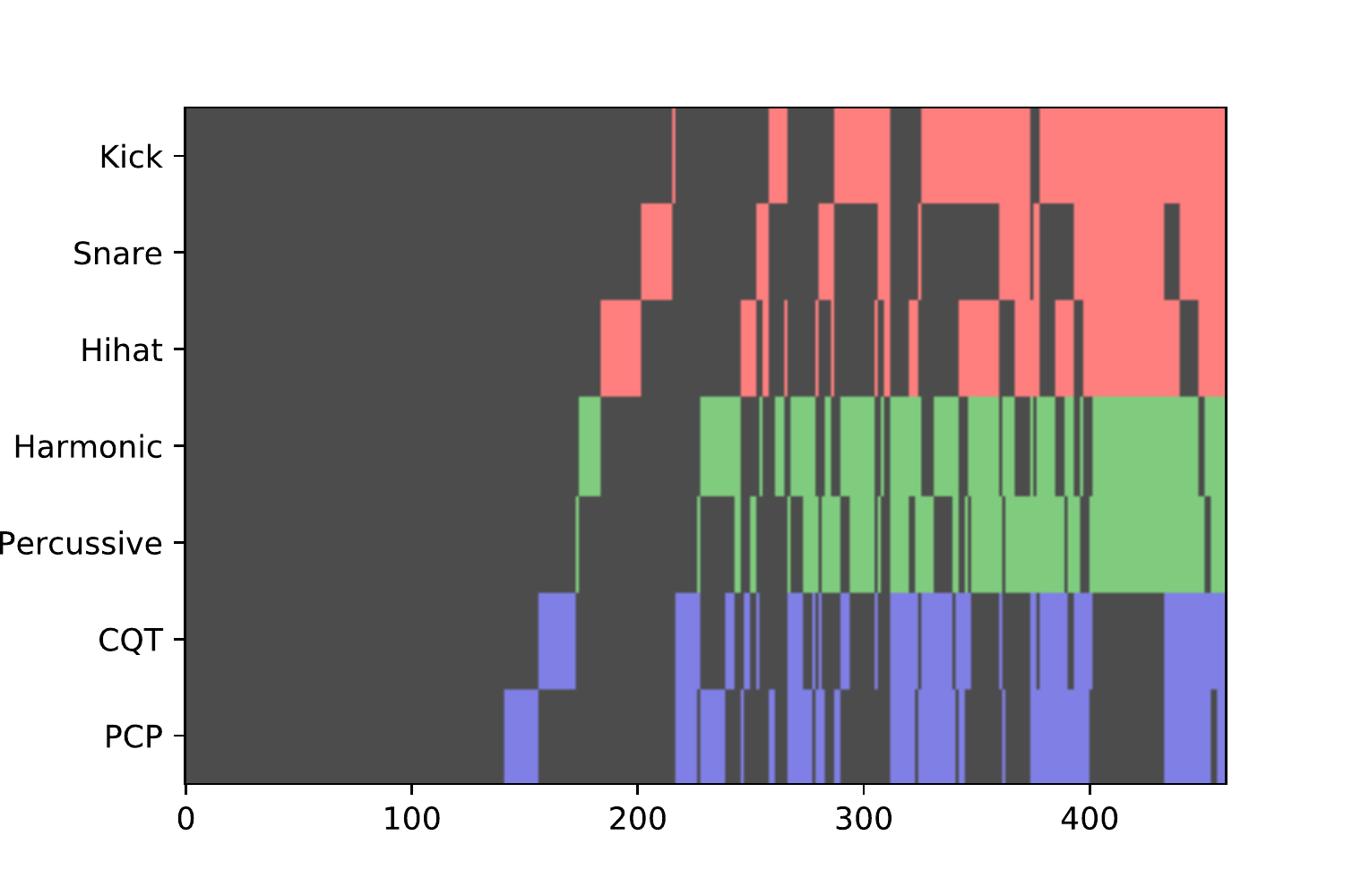
  b)
  \def\svgwidth{\columnwidth}
  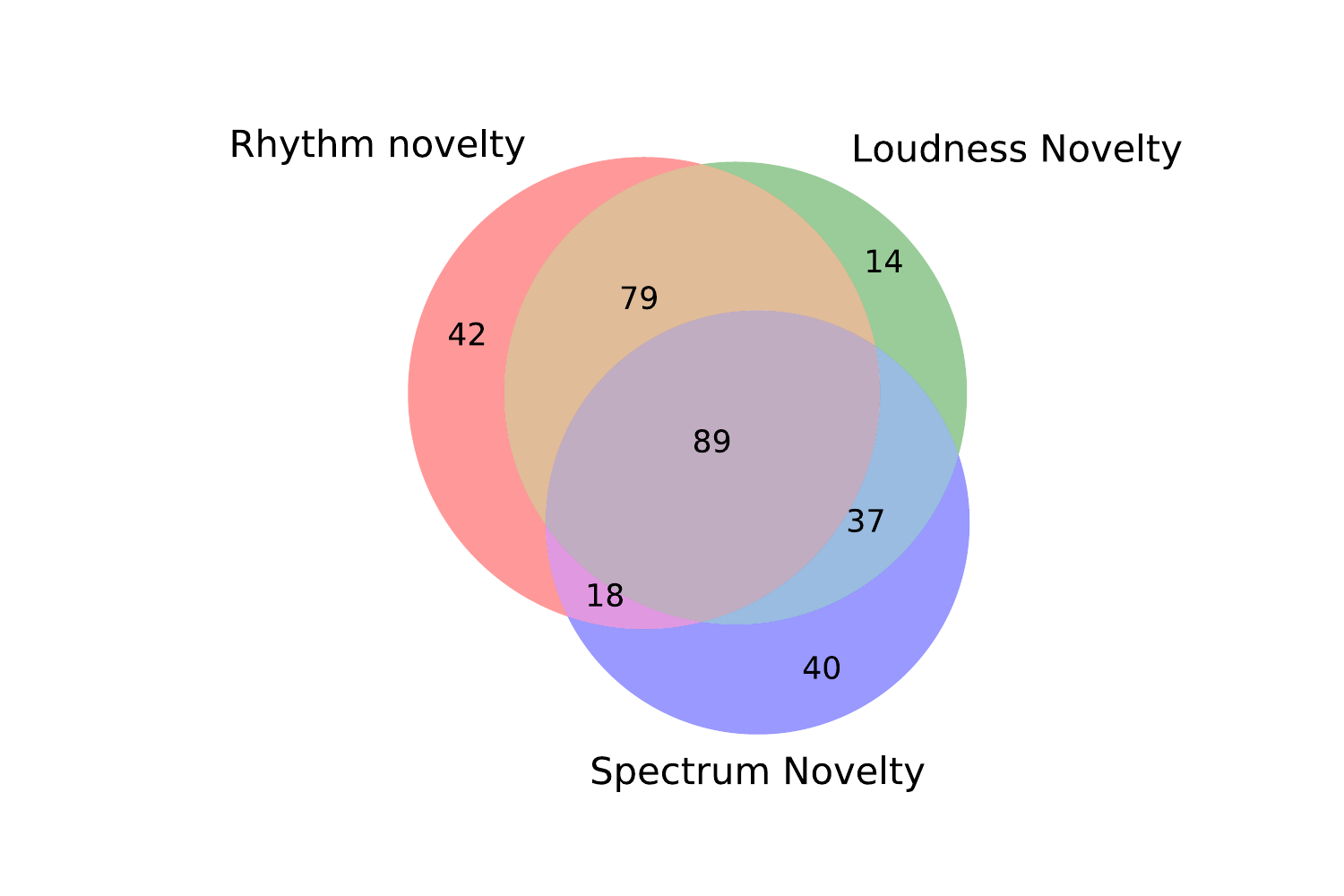
  \caption{Empirical relationship between features. a) Binary heatmap showing when a feature (rows) detects an annotation (columns), with each unit on the x-axis representing one annotation. b) Venn diagram representation of the heatmap. The features are grouped to ease readability and the numbers indicate how many annotations are found by each group or their intersection.}
  \label{fig:featuresCoverage}

\end{figure*}

In Fig.~\ref{fig:featuresCoverage}, we display
the relation between the seven features and the annotations correctly detected by each feature, according to our ground truth dataset and the novelty detection scheme presented below. Even though most of the positions are found by novelty in multiple features at the same time, each category is still contributing independently to the detection of switch points.

\subsection{Novelty detection} \label{sec. novelty detection}

As a preprocessing step to compute novelty, features are aggregated according to a so-called beat-synchronicity (Lines 4--6 of Alg.~\ref{algo:cueDetection}). This step introduces tempo invariance---an advantage that is not available with a frame-synchronous representation---and increases the granularity of the features to a full-time unit of the track.  In our case, we quantized the features to the grid of strong beats (beats 1 and 3 of a 4/4 bar) as this is the coarsest unit of time supported by the Madmom library~\cite{bock_joint_2016} that we could reliably extract from the tracks.\footnote{We found that in EDM it may be problematic to tell apart the downbeat from the 3rd beat of a bar. Because of this, we could not use bar synchronicity reliably.} The quantization was performed by aggregating values of all the samples located between each strong beat. In most cases, features were averaged by computing the Root Mean Square (RMS) value; for the sparse features (concerning drum transcription only), the number of detected onsets was used instead.

The most common approach to compute novelty points in a signal relies on the convolution of a checker-board kernel with the self-similarity matrix built from it (labelled as SSM, Line 7). This method is described in~\cite{foote_automatic_2000} and used in~\cite{vande_veire_raw_2018, davies_automashupper:_2014, lin_audio_2015, kim_automatic_2017}. We used the (strong) beat synchronized representations of each feature to build the associated SSM with the standardized-Euclidean distance. The kernel size was set to compute novelty between segments of four bars (hence, a kernel of eight bars) corresponding to the typical smallest length of a section as described in Sec.~\ref{Rules}. By default, we used a convolution with valid padding which does not return any (trivial) novelty value at the start of a track. For the loudness features only, we found that using the same padding with zero values gives us good results in the intro section of a track.

After extracting the novelty curves, we used a peak-picking algorithm to identify the points of maximal novelty in each feature (Line 8): all of them are potential switch points. The peak-picking procedure works by extracting points which are both a local maximum in a window of four bars and have an amplitude above 0.3 times the maximum value of the feature.

\subsection{Period detection} \label{sec:period detection}
According to Rule \ref{rule:switch point position},
candidates for switch points lie right at the boundaries between sections within a track; since those boundaries are likely to be spaced out by a multiple of four bars, we removed all candidates that do not follow this periodicity (Line 11).
Notice however that due to the presence of anacruses, the first beat of the track is not necessarily the start of the period.
To detect the most probable phase offset of a 4-bar period, we employed a modified version of~\cite{vande_veire_raw_2018}. Specifically, from the eight possible offsets (one for each strong beat candidate), we
select the one that maximizes novelty in the whole track.
To compute the score of each offset, instead of summing and counting only the novelty peaks, we use the RMS of the set of novelties four bars apart.

\subsection{Salience detection}
The candidates identified so far correspond to locations of high novelty for at least one of the seven features which
lie in the musical period. Those locations are good candidates to extract the structural boundaries in the music, but,
to ensure quality in the context of a DJ-mix, with Rule \ref{rule:switch-point prevalent} we impose that the segment following the switch point must be able to stand on its own.
This rule is implemented (Line 12) by thresholding the harmonic content of the track to a minimum for the four bars
following the switch point, thus ensuring that the segment played will be loud (and hopefully salient) enough to stand
out in the mix. We set a threshold at 0.4 times the maximum of the harmonic energy for the average of the four bars
following the switch point. While we fully realize that this implementation is an approximation of the rule since a segment doesn't need to have substantial harmonic energy to be salient, it is restrictive enough to discard all those
candidates that do not satisfy the rule.

\section{Evaluation}
\label{Evaluation}
In the following, we use the terms ``candidates'' and ``annotations'' to denote the switch points returned by
Algorithm~\ref{algo:cueDetection}, and those chosen by music experts, respectively.
In order to evaluate the quality of such candidates,
we curated a dataset of switch points which constitutes our ground truth
(Sec.~\ref{Dataset}); with that, we then assessed the precision achieved by our implementation of the three aforementioned rules, and objectively
compared our approach with other state-of-the-art methods (Sec.~\ref{objective}). Furthermore, aware that the choice of
switch points is a subjective task (ultimately, it is ``matter of taste'') and that no dataset can therefore capture the
ground truth in its entirety, we also conducted a subjective evaluation of those candidates that
do not find a matching annotation in the dataset (Sec.~\ref{subjective}).


\subsection{Dataset}
\label{Dataset}

We created a dataset with annotations for 134 tracks of Electronic Dance Music, selected from a period of 30 years (1987–2016), a variety of musical subgenres, and a tempo ranging from 99bpm to 147bpm.
About 60\% of the tracks come from the digitalization of vinyl records.
All the tracks were converted to a standard compressed format using ffmpeg (128kpbs, 44.1kHz);
the average duration is 7min 25s per track, for a combined duration of 16h 33min.
The dataset was independently annotated by five different musicians with a level of qualification ranging from a professional composer to a semi-professional DJ.

The objective evaluation on this dataset is based on the Hit Rate: A candidate counts as a hit if it is located within a
window of 0.5 seconds from an annotation. We chose this time window as it is typically used for strict evaluation in music structure analysis. With this metric, we are interested in computing the precision (how many candidates are correct), as our objective is to maximize it. For the sake of completeness, we also report the recall (how many annotations are found). 

As mentioned, switch points are chosen subjectively: They depend both on the taste and style of the DJ, and the peculiarities of the track itself.
Because of this, in order to create a homogeneous set of annotations, the annotators were given our set of rules from Sec.~\ref{Rules} as guidelines to follow.
Furthermore, to mitigate the impact of subjectivity, each track was annotated independently by three people and their annotations were merged. This measure increases the ``spread'' of the valid positions. The assignment of each track to three of the five annotators was made according to their availability.

Due to the large number of possible switch points per track, we also required the annotators to constrain the annotations from the start of the track up to the first beat of the core section.
This choice shortens the annotation process while keeping what we believe are the most valuable and commonly used switch points from the intro section (if present) and the start of the first core section.\footnote{It is conceivable to incorporate this constraint also in the algorithm, effectively turning it into a Rule 4.}Accordingly, we only evaluate the candidates (generated by Algorithm~\ref{algo:cueDetection} or the other techniques in Sec. \ref{objective}) which lie in the portion of the track covered by the annotations.


\subsection{Objective Evaluation} \label{objective}

\begin{figure}
  \def\svgwidth{\columnwidth}
  {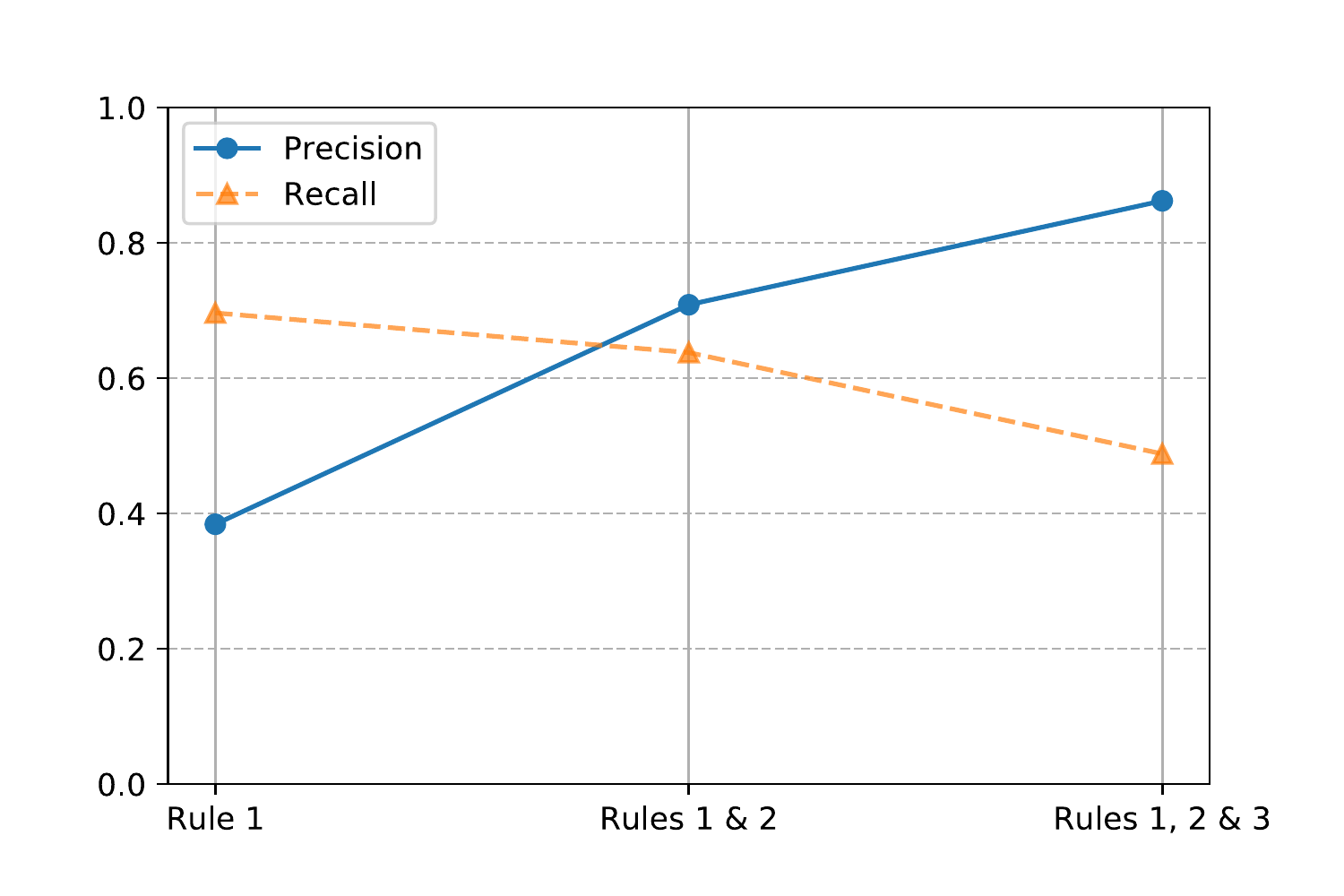}
  \caption{Impact of the different rules on the precision and recall attained by Algorithm~1.}
  \label{fig:resultsPerRule}
\end{figure}

In Fig.~\ref{fig:resultsPerRule}, we evaluate the impact of the three rules, as they were implemented, on the precision achieved by Algorithm 1.
Effectively, Rule 1 is implemented by Lines 1--10, and Lines 11 and 12 correspond to Rules 2 and 3, respectively. Since an automatic DJ-software only needs one switch-in point to progress from one track to the next, precision is the metric we aim to maximize.
Clearly, as more rules are enabled, the algorithm becomes more selective in the choice of candidates, and the precision increases.
With all rules enabled, the precision becomes 86\%, indicating that most of the candidates are of good quality; as the recall is 49\%, the algorithm successfully identifies half of the annotations.


\begin{figure}
  \def\svgwidth{\columnwidth}
  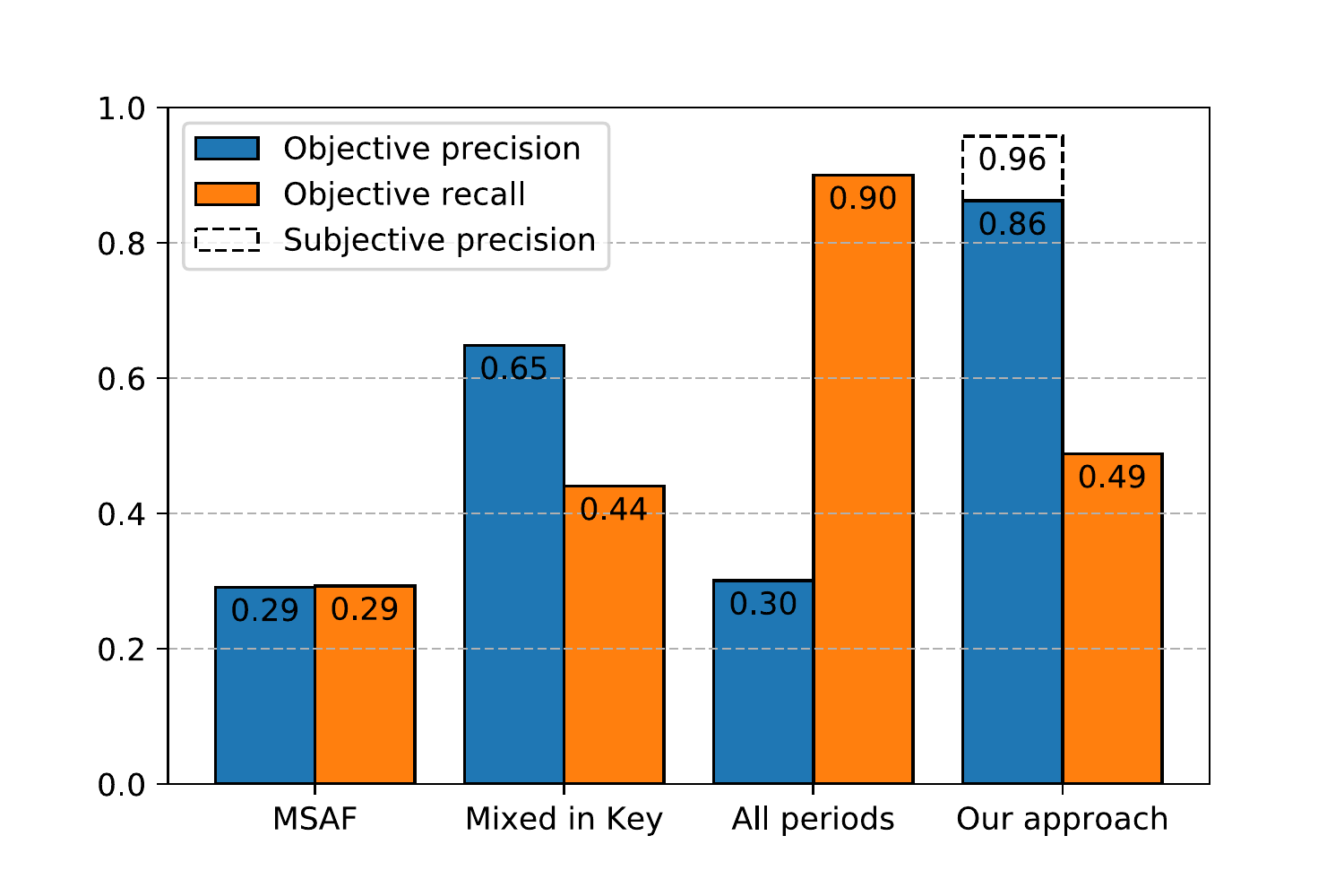
  \caption{Comparison of methods for the generation of switch points.}
  \label{fig:resultsPerAlgo}
\end{figure}

In Fig.~\ref{fig:resultsPerAlgo}, we report on precision and recall of our approach along with that of three other techniques:
\begin{itemize}
  \item ``MSAF:'' From the music structure analysis framework~\cite{nieto_systematic_2016},
        we selected the approach that performs best on our dataset (Algorithm ``OLDA'' with feature
        ``PCP'').
        The structure boundaries so obtained are used as a proxy for switch-in locations.
  \item ``Mixed in Key'' is a commercial product for cue point identification; no algorithmic description of the system is available;
  \item ``All periods'' is an heuristics that selects all the points on the period estimated (see Sec.~\ref{sec:period detection}).
\end{itemize}


Our evaluation does not include the approach described in \cite{vande_veire_raw_2018}, as it only works for music
between 160 to 190 bpm, nor the algorithms in \cite{bittner_automatic_2017,schwarz_heuristic_2018} because the code is not publicly available, and not enough details are provided to reproduce them.

Results in Fig.~\ref{fig:resultsPerAlgo} suggest that for the specific genres and annotations included in our dataset,
our method attains the highest precision ($>85\%$). As expected, ``MSAF'' and the ``All
periods'' heuristics perform badly ($\approx 30\%$).
The former extracts structural boundaries with a fine beat-synchronicity, a task known to be challenging when using a small Hit Rate window~\cite{nieto_systematic_2016};
moreover, the low precision indicates
that structural boundaries alone are not good candidates to identify the actual switch points.
The latter, following Rule \ref{rule:switch point position}, exhaustively lists all possible switch points,
thus achieving a very high recall ($90\%$), but low precision ($30\%$);
the remaining $10\%$ of false negatives (i.e., annotations not matched by any candidate),
do not lie on the estimated period and would thus contrast Rule 2.
In actuality, we determined that those cases are due to
the algorithm wrongly estimating the period (45\% of the false negatives),
tracks that do not follow a strict 4-bar period throughout (36\%),
or incorrect annotations (18\%).
Finally, while the precision achieved by ``Mixed in Key'' is $65\%$, this software does not necessarily aim to identify
switch points according to our set of rules, so no conclusion should be drawn regarding all those candidate switch
points that do not find a match in our dataset.


\subsection{Subjective Evaluation} \label{subjective}
Aware that transitions can take one of many forms, we feel that it is preferable to evaluate false
positives for their pertinence as potential candidates instead of discarding them \textit{a priori} for not being
present in the ground truth. This test
consisted in asking our experts to listen to each of the false positives and to judge whether or not they are suitable
to be a switch point. Table \ref{tab:qualitativeFalsePositive} summarises the
results.
The majority (69.4\%) of the false positives were identified as suitable for a mix by at least one expert. The remaining candidates are truly false positive, either caused by a wrong beat estimation (8.3\%), a wrong period estimation (5.6\%), or not containing sufficient novelty (16.6\%).
By combining the results of both evaluations, 95.8\% of the candidates generated by our algorithm are usable switch points.
On average, 8.9 candidates were generated for each track (std: 5.3, min: 1, max: 33).

\begin{table}
  \bigskip
  \begin{center}
    \begin{tabular}{llllll}
      \toprule
      Reason                               & \#Points & Percent \\
      \midrule
      Beat detected is off                 & 3        & 8.3\%   \\
      Period estimated is off              & 2        & 5.6\%   \\
      No perceivable novelty               & 6        & 16.6\%  \\
      Judged useful by at least one expert & 25       & 69.4\%  \\
      \bottomrule
    \end{tabular}
  \end{center}
  \caption{Individual evaluation of false positive.}
  \label{tab:qualitativeFalsePositive}
\end{table}



\section {Conclusions} \label{Conclusions}
As part of our pursuit for automatic DJ mixing of electronic dance music, we considered the problem of detecting switch
points. These points identify the location in time at which a DJ would transition from one track to the next and are
often referred to as ``cue points''. The approach we developed is motivated by the rather rigid and repetitive structure
of EDM's tracks and is inspired by the actual thought process that DJs follow in live performances. We encoded this
knowledge into a set of rules and used them as guidelines to identify possible candidates for switch points.

In general, our approach consists in identifying boundaries in the structure of a track that correspond to points of a
high novelty for a listener, and in picking those that can be useful for a DJ. Both the construction and the selection
of such boundaries is rule-driven.

To enable fast and objective assessment of different approaches for the identification of switch points,
we created a dataset of 134 tracks, each annotated by several experts.
We hope that this dataset, the only one currently publicly available to our knowledge, will help conduct evaluations on
a greater scale than what has been done hitherto. Far from covering the almost infinite possible solutions for the task,
this approach significantly reduced the amount of work needed for a subjective evaluation, as we had to do as a final
check for our method. It also proved that most of those candidates that were not annotated (false positives) are still
of sufficient quality to be used in a mix.

In summary, both the attained precision and the number of candidates generated are a strong indication that our
algorithm is well suited for the ultimate goal of automatic mixing.



\bibliography{bibliography}

\end{document}